\documentclass{nature_withfigs}

\usepackage{amsmath,amsfonts,amssymb}
\usepackage{graphicx}
\usepackage[ansinew]{inputenc}

\newcommand{\ket}[1]{\ensuremath{| #1\rangle}}   

\title{Towards fault-tolerant quantum computing with trapped ions}

\author{J.~Benhelm$^{1,2}$, G.~Kirchmair$^{1,2}$, C.~F.~Roos$^{1,2}$ \& R.~Blatt$^{1,2}$}

\begin{document}

\maketitle

\begin{affiliations}
\item Institut f{\"u}r Experimentalphysik, Universit{\"a}t
Innsbruck, Technikerstr.~25, A-6020 Innsbruck, Austria \item
Institut f{\"u}r Quantenoptik und Quanteninformation,
{\"O}sterreichische Akademie der Wissenschaften,
Otto-Hittmair-Platz~1, A-6020 Innsbruck, Austria
\end{affiliations}

\begin{abstract}
Today ion traps are among the most promising physical systems for
constructing a quantum device harnessing the computing power
inherent in the laws of quantum
physics\cite{Nielsen2000,Cirac1995}. The standard circuit model of
quantum computing requires a universal set of quantum logic gates
for the implementation of arbitrary quantum operations. As in
classical models of computation, quantum error correction
techniques\cite{Shor1995,Steane1996} enable rectification of small
imperfections in gate operations, thus allowing for perfect
computation in the presence of noise. For fault-tolerant
computation\cite{Shor1996}, it is commonly believed that error
thresholds ranging between $10^{-4}$ and $10^{-2}$
\cite{Knill2005,Raussendorf2007,Reichardt2004} will be required
depending on the noise model and the computational overhead for
realizing the quantum gates. Up to now, all experimental
implementations have fallen short of these requirements. Here, we
report on a
M{\o}lmer-S{\o}rensen\cite{Soerensen1999,Soerensen2000} type gate
operation entangling ions with a fidelity of 99.3(1)\% which
together with single-qubit operations forms a universal set of
quantum gates. The gate operation is performed on a pair of qubits
encoded in two trapped calcium ions using a single
amplitude-modulated laser beam interacting with both ions at the
same time. A robust gate operation, mapping separable states onto
maximally entangled states is achieved by adiabatically switching
the laser-ion coupling on and off. We analyse the performance of a
single gate and concatenations of up to 21 gate operations. The
gate mechanism holds great promise not only for two-qubit but also
for multi-qubit operations.
\end{abstract}

For ion traps, all building blocks necessary for the construction
of a universal quantum computer\cite{Nielsen2000} have been
demonstrated over the last decade. Currently, the most important
challenges consist in scaling up the present systems to higher
number of qubits and in raising the fidelity of gate operations up
to the point where quantum error correction techniques can be
successfully applied. While single-qubit gates are easily
performed with high quality, the realisation of high-fidelity
entangling two-qubit
gates\cite{Sackett2000,Schmidt-Kaler2003,Leibfried2003a,Haljan2005b,Home2006,Riebe:2006a}
is much more demanding since the inter-ion distance is orders of
magnitude bigger than the characteristic length scale of any
state-dependent ion-ion interaction. Apart from quantum gates of
the Cirac-Zoller type\cite{Cirac1995,Schmidt-Kaler2003} where a
laser couples a single qubit with a vibrational mode of the ion
string at a time, most other gate realizations entangling ions
have relied on collective interactions of the qubits with the
laser control
fields\cite{Sackett2000,Leibfried2003a,Haljan2005b,Home2006}.
These gate operations entangle transiently the collective
pseudo-spin of the qubits with the vibrational mode and produce
either a conditional phase shift\cite{Milburn2000} or a collective
spin flip\cite{Soerensen1999,Soerensen2000,Solano1999} of the
qubits. While the highest fidelity $F=97\%$ reported to
date\cite{Leibfried2003a} has been achieved with a conditional
phase gate acting on a pair of hyperfine qubits in $^9$Be$^+$,
spin flip gates have been limited so far to
$F\approx85\%$\cite{Haljan2005b,Sackett2000}. All of these
experiments have used qubits encoded in hyperfine or Zeeman ground
states and a Raman transition mediated by an electric-dipole
transition for coupling the qubits. While spontaneous scattering
from the mediating short-lived levels degrades the gate fidelity
due to the limited amount of laser power available in current
experiments, this source of decoherence does not occur for optical
qubits, i.e. qubits encoded in a ground state and a metastable
electronic state of an ion. In the experiment presented in this
paper where the qubit is comprised of the states $|S\rangle\equiv
S_{1/2}(m=1/2)$ and $|D\rangle\equiv D_{5/2}(m=3/2)$ of the
isotope $^{40}$Ca$^+$, spontaneous decay of the metastable state
reduces the gate fidelity by less than $5\cdot 10^{-5}$.

\begin{figure}
\begin{center}
\includegraphics[width=89mm]{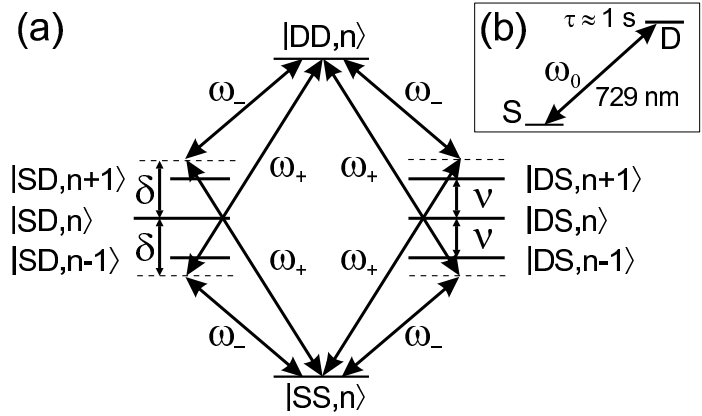}
\caption{\label{gatescheme} \textbf{Gate mechanism.} (a) A
bichromatic laser field with frequencies $\omega_+$, $\omega_-$
satisfying 2$\omega_0$ = $\omega_+$ + $\omega_-$ is tuned close to
the upper and lower motional sideband of the qubit transition. The
field couples the qubit states \ket{SS} $\leftrightarrow$ \ket{DD}
via the four interfering paths shown in the figure, $n$ denoting
the vibrational quantum number of the axial COM mode. Similar
processes couple the states $\ket{SD}\leftrightarrow\ket{DS}$. (b)
The qubits are encoded in the ground state $S_{1/2}(m=1/2)$ and
the metastable state $D_{5/2}(m=3/2)$ of $^{40}$Ca$^+$ ions and
manipulated by a narrow bandwidth laser emitting at a wavelength
of 729~nm.}
\end{center}
\end{figure}

A M{\o}lmer-S{\o}rensen gate inducing collective spin flips is
achieved with a bichromatic laser field with frequencies
$\omega_\pm=\omega_0\pm\delta$, with $\omega_0$ the qubit
transition frequency and $\delta$ close to the vibrational mode
frequency $\nu$ (see Fig.~\ref{gatescheme}). For optical qubits,
the bichromatic field can be a pair of co-propagating lasers which
is equivalent to a single laser beam resonant with the qubit
transition and amplitude-modulated with frequency $\delta$. For a
gate mediated by the axial centre-of-mass (COM) mode, the
Hamiltonian describing the laser-qubit interaction is given by $H
= \hbar\Omega e^{-i\phi}S_+ (e^{-i(\delta t+\zeta)}+e^{i(\delta
t+\zeta)})e^{i\eta(ae^{-i\nu t}+a^\dagger e^{i\nu t})} +
\mbox{h.c.}$. Here,
$S_j=\sigma_j^{(1)}+\sigma_j^{(2)},j\in\{+,-,x,y,z\}$, denotes a
collective atomic operator, and
$\sigma_+^{(i)}|S\rangle_i=|D\rangle_i$. The operators
$a$,~$a^\dagger$ annihilate and create phonons of the COM mode
with Lamb-Dicke factor $\eta$. The optical phase of the laser
field is labeled $\phi$, and the phase $\zeta$ accounts for a time
difference between the start of the gate operation and the maximum
of the amplitude modulation of the laser beam. In the Lamb-Dicke
regime, and for $\phi=0$, the gate operation is very well
described by the propagator\cite{Roos2007}
\begin{equation}
U(t)=e^{-iF(t)S_x}\hat{D}(\alpha(t)S_{y,\psi})\exp(-i(\lambda
t+\chi\sin(\nu\!-\!\delta)t)S_{y,\psi}^{\;2}). \label{propagator}
\end{equation}
Here, the operator to the right describes collective spin flips
induced by the operator $S_{y,\psi}=S_y\cos\psi+S_z\sin\psi$,
$\psi=\frac{4\Omega}{\delta}\cos\zeta$, and
$\lambda\approx\eta^2\Omega^2/(\nu-\delta)$,
$\chi\approx\eta^2\Omega^2/(\nu-\delta)^2$. With
$\alpha(t)=\alpha_0(e^{i(\nu-\delta)t}-1)$, the displacement
operator $\hat{D}(\beta)=e^{\beta a^\dagger-\beta^\ast a}$
accounting for the transient entanglement between the qubits and
the harmonic oscillator becomes equal to the identity after the
gate time $\tau_{gate}=\frac{2\pi}{|\nu-\delta|}$. The operator
$e^{-iF(t)S_x}$ with
$F(t)=(2\Omega/\delta)(\sin(\delta t+\phi)-\sin\phi)$
describes fast non-resonant excitations of the carrier transition
that occur in the limit of short gates when $\Omega\ll\delta$ no
longer strictly holds. Non-resonant excitations are suppressed by
intensity-shaping the laser pulse so that the Rabi frequency
$\Omega(t)$ is switched on and off smoothly. Moreover, adiabatic
switching makes the collective spin flip operator independent of
$\zeta$ as $S_{y,\psi}\rightarrow S_y$ for $\Omega\rightarrow 0$.
To achieve adiabatic following, it turns out to be sufficient to
switch on the laser within 2.5 trap cycles. In order to realize an
entangling gate of duration $\tau_{gate}$ described by the unitary
operator $U_{gate}=\exp(-i\frac{\pi}{8}S_y^2)$, the laser
intensity needs to be set such that $\eta\Omega\approx
|\delta-\nu|/4$.

Two $^{40}$Ca$^+$ ions are confined in a linear
trap\cite{Benhelm2007} with axial and radial frequencies of
$\nu_{axial}/2\pi = 1.23$~MHz and $\nu_{radial}/2\pi = 4$~MHz,
respectively. After Doppler cooling and frequency-resolved optical
pumping\cite{Roos2006}, the two axial modes are cooled close to
the motional ground state ($\bar{\mbox{n}}_{\mbox{\footnotesize
com}}$, $\bar{\mbox{n}}_{\mbox{\footnotesize stretch}} <$
0.05(5)). Both ions are now initialized to \ket{SS} with a
probability of more than 99.8\%. Then, the gate operation is
performed, followed by an optional carrier pulse for analysis.
Finally,  we measure the probability $p_k$ of finding $k$ ions in
the $|S\rangle$ state by detecting light scattered on the $S_{1/2}
\leftrightarrow~P_{1/2}$ dipole transition with a photomultiplier
for 3~ms. The error in state detection due to spontaneous emission
is estimated to be less than 0.15\%. Each experimental cycle is
synchronised with the frequency of the AC-power line and repeated
50-200 times. The laser beam performing the entangling operation
is controlled by a double-pass acousto-optic modulator (AOM) which
allows setting the frequency $\omega_L$ and phase $\phi$ of the
beam. By means of a variable gain amplifier, we control the
radio-frequency (r.f.) input power and hence the intensity profile
of each laser pulse. To generate a bichromatic light field, the
beam is passed through another AOM in single-pass configuration
that is driven simultaneously by two r.f. signals with difference
frequency $\delta/\pi$. Phase coherence of the laser frequencies
is maintained by phase-locking all r.f. sources to an ultra-stable
quartz oscillator. We use 1.8~mW of light focused down to a spot
size of 14~$\mu$m Gaussian beam waist illuminating both ions from
an angle of 45$°$ with equal intensity to achieve the Rabi
frequencies $\Omega/(2\pi)\approx 110$~kHz required for performing
a gate operation with $(\nu-\delta)/(2\pi)=20$~kHz and
$\eta=0.044$. To make the bichromatic laser pulses independent of
the phase $\zeta$, the pulse is switched on and off by using pulse
slopes of duration $\tau_r$ = 2~$\mu$s.
%

\begin{figure}
\includegraphics[width=183mm]{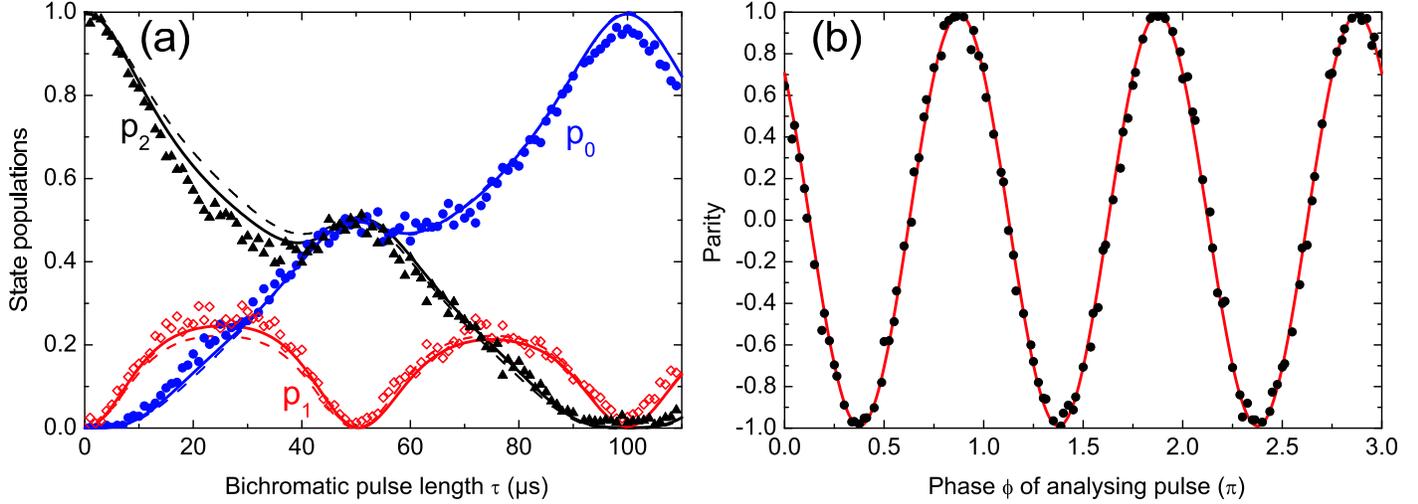}
\caption{\label{timeevolution}\label{onegate} \textbf{High-fidelity
gate operation.} (a) State evolution induced by a
M{\o}lmer-S{\o}rensen bichromatic pulse of duration $\tau$. The Rabi
frequency $\Omega(t)$ is smoothly switched on and off within $2~\mu
s$ and adjusted such that a maximally entangled state is created at
$\tau_{gate}=50~\mu s$. The dashed lines are calculated for
$\bar{\mbox{n}}_{\mbox{\footnotesize com}} = 0.05$ from the
propagator (\ref{propagator}), neglecting pulse shaping and
non-resonant carrier excitation. The solid lines are obtained from
numerically solving the Schr\"odinger equation for time-dependent
$\Omega(t)$ and imbalanced Rabi frequencies
$\Omega_{+}/\Omega_{-}=1.094$
(see Methods). (b) A $(\frac{\pi}{2})_\phi$ analysis pulse applied
to both ions prepared in $\Psi_1$ gives rise to a parity oscillation
$P(\phi)=\sin(2\phi)$ as a function of $\phi$. A fit with a function
$P_{fit}=A\sin(2\phi+\phi_0)$ yields the parity oscillation
amplitude $A=0.990(1)$ and  $\phi_0/\pi=-1.253(1)$. The precise
value of the phase $\phi_0$ is without significance. It arises from
phase-locking the frequencies $\omega_0, \omega_+, \omega_-$ and
could have been experimentally adjusted to zero.}
\end{figure}

Multiple application of the bichromatic pulse of duration
$\tau_{gate}$ ideally maps the state $|SS\rangle$ to
\begin{equation}
\label{evolution} \ket{SS} \stackrel{\tau_{gate}}{\longrightarrow}
\underbrace{\ket{SS}+i\ket{DD}}_{\Psi_1}
\stackrel{\tau_{gate}}{\longrightarrow} \ket{DD}
\stackrel{\tau_{gate}}{\longrightarrow}\ket{DD}+i\ket{SS}
 \stackrel{\tau_{gate}}{\longrightarrow}\ket{SS}
\stackrel{\tau_{gate}}{\longrightarrow}...
\end{equation}
up to global phases. Maximally entangled states occur at instances
$\tau_m = m\cdot\tau_{gate}$ ($m=1,3,\ldots$). A similar mapping
of product states onto Bell states and vice versa also occurs when
starting from state $|SD\rangle$. In order to assess the fidelity
of the gate operation, we adapt the strategy first applied in
refs.\cite{Sackett2000,Leibfried2003a} consisting in measuring the
fidelity of Bell states created by a single application of the
gate to the state $|SS\rangle$ (see Fig.~\ref{onegate} (a)). The
fidelity
$F=\langle\Psi_1|\rho^{exp}|\Psi_1\rangle=(\rho^{exp}_{SS,SS}+\rho^{exp}_{DD,DD})/2+\mbox{Im}\rho^{exp}_{DD,SS}$,
with the density matrix $\rho^{exp}$ describing the experimentally
produced qubits' state, is inferred from measurements on a set of
$42,400$ Bell states continuously produced within a measurement
time of $35$ minutes. Fluorescence measurements on $13,000$ Bell
states reveal that
$\rho^{exp}_{SS,SS}+\rho^{exp}_{DD,DD}=p_2+p_0=0.9965(4)$. The
off-diagonal element $\rho^{exp}_{DD,SS}$ is determined by
measuring
$P(\phi)=\langle\sigma_\phi^{(1)}\sigma_\phi^{(2)}\rangle$ for
different values of $\phi$, where
$\sigma_\phi=\sigma_x\cos\phi+\sigma_y\sin\phi$, by applying
$(\frac{\pi}{2})_\phi$-pulses to the remaining $29,400$ states and
measuring $p_0+p_2-p_1$ to obtain the parity
$\langle\sigma_z^{(1)}\sigma_z^{(2)}\rangle$. The resulting parity
oscillation $P(\phi)$ shown in Fig.~\ref{onegate} (b) is fitted
with a function $P_{fit}(\phi)=A\sin(2\phi+\phi_0)$ that yields
$A=2|\rho^{exp}_{DD,SS}|=0.990(1)$. Combining the two
measurements, we obtain the fidelity $F=99.3(1)\%$ for the Bell
state $\Psi_1$.

\begin{figure}
\includegraphics[width=183mm]{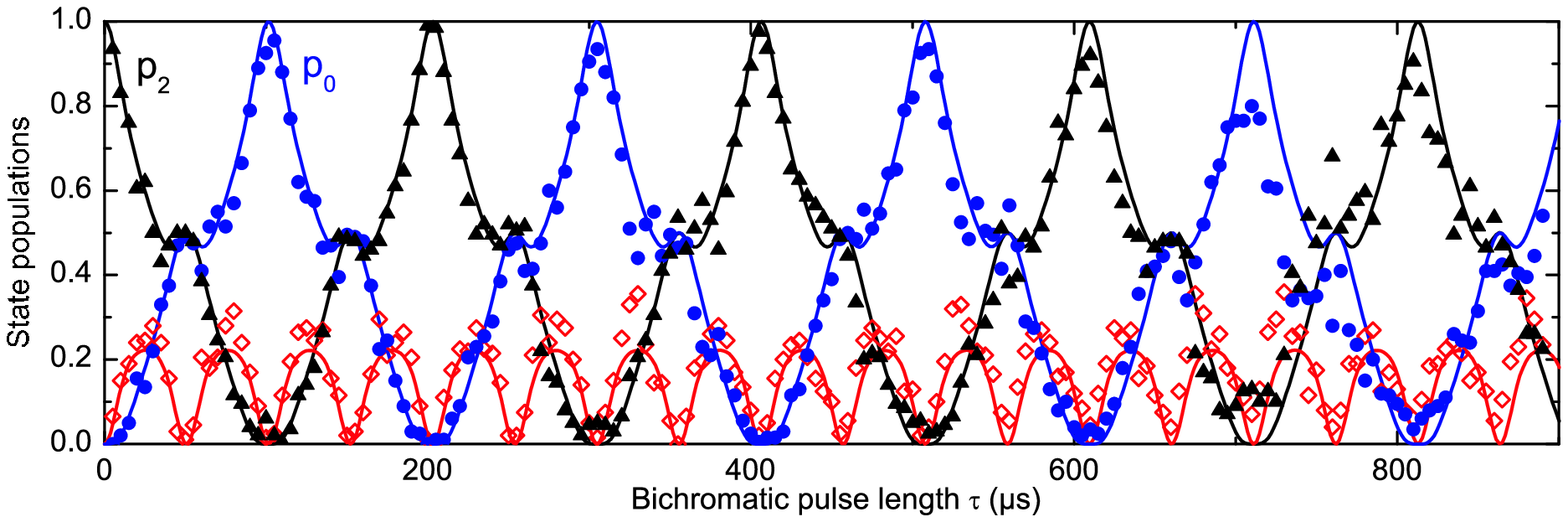}
\caption{\label{manygates} \textbf{Entanglement and
disentanglement dynamics of the M{\o}lmer-S{\o}rensen
interaction.} Starting from state $|SS\rangle$ for a detuning of
the bichromatic laser from the sidebands set to
$\delta-\nu=-20$~kHz, the figure shows the time evolution of the
populations $p_0$, $p_1$, and $p_2$ denoted by the symbols
($\bullet$),
($\diamond$), and ($\blacktriangle$), respectively. The length of
the pulse is equivalent to the application of up to 17 gate
operations. Maximally entangled states are created whenever
$p_0(\tau)$ and $p_2(\tau)$ coincide and $p_1(\tau)$ vanishes.}
\end{figure}

A wealth of further information is obtained by studying the state
dynamics under the action of the gate Hamiltonian. Starting from
state $|SS\rangle$, Fig.~\ref{manygates} depicts the time evolution
of the state populations for pulse lengths equivalent up to 17 gate
times. The ions are entangled and disentangled consecutively up to
nine times, the populations closely following the predicted unitary
evolution of the propagator (\ref{propagator}) for $\zeta=0$ shown
in Fig.~\ref{manygates} as a solid line.

\begin{figure}
\begin{center}
\includegraphics[width=89mm]{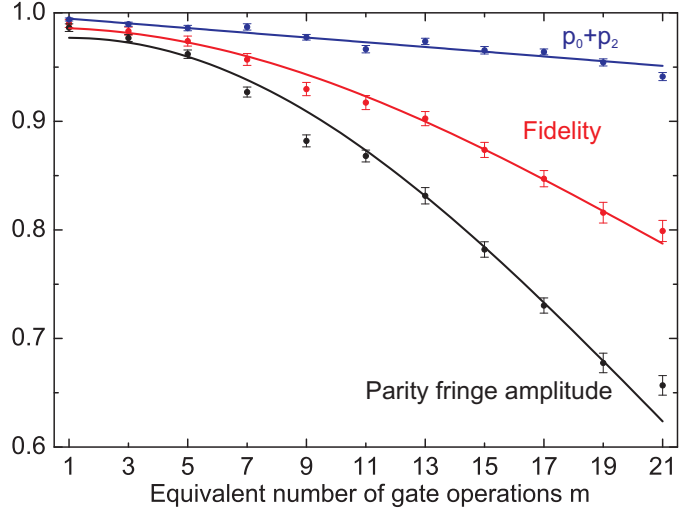}
\caption{\label{MultiGateFidelity} \textbf{Multiple gate
operations.} Gate imperfections as a function of the bichromatic
pulse length $\tau_m=m\cdot\tau_{gate}$ given in equivalent number
of gate operations $m$. The upper curve shows a linear decrease of
the state populations $p_0+p_2$ with a slope of 0.0022(1). The
lower curve the magnitude of the coherence $2\rho_{DD,SS}$
measured by detecting parity oscillations and fitted by a Gaussian
decay function that accounts for low-frequency noise of the laser
frequency and the magnetic field. Combining both measurements
yields the Bell state fidelity $F_m$ shown as the middle trace.
For $m=21$, the fidelity is still $F_{21}=80(1)\%$. Similar
results are achieved when replacing the entangling pulse of length
$\tau_m$ by $m$ amplitude-shaped pulses each of which is realizing
an entangling gate operation.}
\end{center}
\end{figure}

To study sources of gate imperfections we measured the fidelity of
Bell states obtained after a pulse length $\tau_m$ for up to m =
21 gate operations. The sum of the populations $p_0(t)+p_2(t)$
does not return perfectly to one at times $\tau_{m}$ as shown in
Fig.~\ref{MultiGateFidelity} but decreases by about 0.0022(1) per
gate. This linear decrease could be explained by resonant spin
flip processes caused by spectral components of the qubit laser
that are far outside the laser's linewidth of 20~Hz (see Methods).
The figure also shows the amplitude of parity oscillation scans at
odd integer multiples of $\tau_{gate}$ similar to the one in
Fig.~\ref{onegate} (b). The Gaussian shape of the amplitude decay
is consistent with low-frequency noise of the magnetic field and
the laser frequency as the source of imperfections (see Methods).


The observed Bell state infidelity of $7\cdot10^{-3}$ suggests
that the gate operation has an infidelity below the error
threshold required by some models of fault-tolerant quantum
computation\cite{Knill2005,Raussendorf2007,Reichardt2004}.
However, further experimental advances will be needed before
fault-tolerant computation will become a reality as the overhead
implied by these models is considerable. Nevertheless, in addition
to making the implementation of quantum algorithms with tens of
entangling operations look realistic, the gate presented here also
opens interesting perspectives for generating multi-particle
entanglement\cite{Moelmer1999} by a single laser interacting with
more than two qubits at once. For the generation of N-qubit GHZ
states, there exist no constraints on the positioning of ions in
the bichromatic beam that otherwise makes generation of GHZ states
beyond N=6 so difficult in the case of hyperfine
qubits\cite{Leibfried2005}. While the bichromatic force lacks a
strong spatial modulation that would enable tailoring of the gate
interaction by choosing particular ion
spacings\cite{Chiaverini2004,Reichle2006a}, more complex
multi-qubit interactions could be engineered by interleaving
entangling laser pulses addressing all qubits with a focussed
laser inducing phase shifts in single qubits. Akin to nuclear
magnetic resonance techniques, this method should allow for
refocussing of unwanted qubit-qubit
interactions\cite{Vandersypen2004} and open the door to a wide
variety of entangling multi-qubit interactions.
%


\newpage
\begin{methods}
\subsection{AC-Stark-shift compensation}
The red- and the blue-detuned frequency components $\omega_\pm$ of
the bichromatic light field cause dynamic (ac-) Stark shifts by
non-resonant excitation on the carrier and the first-order
sidebands that exactly cancel each other if the corresponding
laser intensities $I_\pm$ are equal. The remaining ac-Stark shift
due to other Zeeman transitions and far-detuned dipole transitions
amounts to 7~kHz for a gate time $\tau_{gate}=50~\mu$s. These
shifts could be compensated by using an additional far-detuned
light field\cite{Haeffner2003} or by properly setting the
intensity ratio $I_{+}/I_{-}$. We utilize the latter technique
which makes the coupling strengths $\Omega_{SS\leftrightarrow
DD}\propto 2\sqrt{I_{+}I_{-}}$, $\Omega_{SD\leftrightarrow
DS}\propto I_{+}+I_{-}$ slightly unequal. However, the error is
insignificant as $\Omega_{SD\leftrightarrow
DS}/\Omega_{SS\leftrightarrow DD}-1=4\cdot 10^{-3}$ in our
experiments.

\subsection{Sources of gate infidelity}
Spin flips induced by incoherent off-resonant light of the
bichromatic laser field reduce the gate fidelity. A beat frequency
measurement between the gate laser and a similar independent laser
system that was spectrally filtered indicates that a fraction
$\gamma$ of about $2\cdot 10^{-7}$ of the total laser power is
contained in a 20~kHz bandwidth $B$ around the carrier transition
when the laser is tuned close to a motional sideband. A simple model
predicts spin flips to cause a gate error with probability
$p_{flip}=(\pi\gamma|\nu-\delta|)/(2\eta^2B)$. This would correspond
to a probability $p_{flip}=8\cdot 10^{-4}$ whereas the measured
state populations shown in Fig.~\ref{MultiGateFidelity} would be
consistent with $p_{flip}=2\cdot 10^{-3}$. Spin flip errors could be
further reduced by two orders of magnitude by spectrally filtering
the laser light and increasing the trap frequency $\nu$ to above
2~MHz where noise caused by the laser frequency stabilization is
much reduced. Variations in the coupling strength $\Omega$ induced
by low-frequency laser intensity noise and thermally occupied radial
modes can be independently measured by recording the amplitude decay
of carrier oscillations. For $\delta\Omega/\Omega$, we find a
variation of $7\cdot 10^{-3}$. The Gaussian decay of the parity
contrast shown in Fig. (\ref{MultiGateFidelity}) is attributed to
low frequency noise randomly shifting the laser frequency with
respect to the atomic transition frequency $\omega_0$ and having a
Gaussian distribution with full-width at half maximum of
$\delta\omega/(2\pi)=180$~Hz. This value is consistent with Ramsey
measurements on a single ion predicting
$\delta\omega/(2\pi)=160$~Hz. For a single gate, the frequency
uncertainty gives rise to a fidelity loss of 0.1\%.

A bichromatic force with time-dependent $\Omega(t)$ acting on ions
prepared in an eigenstate of $S_y$ creates coherent states
$\alpha(t)$ following trajectories in phase space that generally
do not close\cite{Leibfried2007,Roos2007}. For the short rise
times used in our experiments, this effect can be made negligibly
small by slightly increasing the gate time.

\end{methods}

\newpage

Bibliography
\bibliographystyle{naturemag}
\bibliography{paper}

\begin{thebibliography}{10}
\expandafter\ifx\csname url\endcsname\relax
  \def\url#1{\texttt{#1}}\fi
\expandafter\ifx\csname urlprefix\endcsname\relax\def\urlprefix{URL }\fi
\providecommand{\bibinfo}[2]{#2}
\providecommand{\eprint}[2][]{\url{#2}}

\bibitem{Nielsen2000}
\bibinfo{author}{Nielsen, M.~A.} \& \bibinfo{author}{Chuang, I.~L.}
\newblock \emph{\bibinfo{title}{Quantum Computation and Quantum Information}}
  (\bibinfo{publisher}{Cambridge Univ. Press, Cambridge},
  \bibinfo{year}{2000}).

\bibitem{Cirac1995}
\bibinfo{author}{Cirac, J.~I.} \& \bibinfo{author}{Zoller, P.}
\newblock \bibinfo{title}{Quantum computations with cold trapped ions.}
\newblock \emph{\bibinfo{journal}{Phys. Rev. Lett.}}
  \textbf{\bibinfo{volume}{74}}, \bibinfo{pages}{4091--4094}
  (\bibinfo{year}{1995}).

\bibitem{Shor1995}
\bibinfo{author}{Shor, P.~W.}
\newblock \bibinfo{title}{Scheme for reducing decoherence in quantum computer
  memory.}
\newblock \emph{\bibinfo{journal}{Phys. Rev. A}} \textbf{\bibinfo{volume}{52}},
  \bibinfo{pages}{R2493--R2496} (\bibinfo{year}{1995}).

\bibitem{Steane1996}
\bibinfo{author}{Steane, A.~M.}
\newblock \bibinfo{title}{Error correcting codes in quantum theory.}
\newblock \emph{\bibinfo{journal}{Phys. Rev. Lett.}}
  \textbf{\bibinfo{volume}{77}}, \bibinfo{pages}{793--797}
  (\bibinfo{year}{1996}).

\bibitem{Shor1996}
\bibinfo{author}{Shor, P.~W.}
\newblock \bibinfo{title}{Fault-tolerant quantum computation}.
\newblock In \emph{\bibinfo{booktitle}{37th Symposium on Foundations of
  Computing}}, \bibinfo{pages}{56--65} (\bibinfo{publisher}{IEEE Computer
  Society Press}, \bibinfo{year}{1996}).

\bibitem{Knill2005}
\bibinfo{author}{Knill, E.}
\newblock \bibinfo{title}{Quantum computing with realistically noisy devices}.
\newblock \emph{\bibinfo{journal}{Nature}} \textbf{\bibinfo{volume}{434}},
  \bibinfo{pages}{39--44} (\bibinfo{year}{2005}).

\bibitem{Raussendorf2007}
\bibinfo{author}{Raussendorf, R.} \& \bibinfo{author}{Harrington, J.}
\newblock \bibinfo{title}{Fault-tolerant quantum computation with high
  threshold in two dimensions}.
\newblock \emph{\bibinfo{journal}{Physical Review Letters}}
  \textbf{\bibinfo{volume}{98}}, \bibinfo{pages}{190504}
  (\bibinfo{year}{2007}).

\bibitem{Reichardt2004}
\bibinfo{author}{Reichardt, B.~W.}
\newblock \bibinfo{title}{Improved ancilla preparation scheme increases
  fault-tolerant threshold}.
\newblock \emph{\bibinfo{journal}{arXiv:quant-ph/0406025v1}} \bibinfo{pages}{4}
  (\bibinfo{year}{2004}).

\bibitem{Soerensen1999}
\bibinfo{author}{S{\o}rensen, A.} \& \bibinfo{author}{M{\o}lmer, K.}
\newblock \bibinfo{title}{Quantum computation with ions in thermal motion}.
\newblock \emph{\bibinfo{journal}{Phys. Rev. Lett.}}
  \textbf{\bibinfo{volume}{82}}, \bibinfo{pages}{1971--1974}
  (\bibinfo{year}{1999}).

\bibitem{Soerensen2000}
\bibinfo{author}{S{\o}rensen, A.} \& \bibinfo{author}{M{\o}lmer, K.}
\newblock \bibinfo{title}{Entanglement and quantum computation with ions in
  thermal motion}.
\newblock \emph{\bibinfo{journal}{Phys. Rev. A}} \textbf{\bibinfo{volume}{62}},
  \bibinfo{pages}{022311} (\bibinfo{year}{2000}).

\bibitem{Sackett2000}
\bibinfo{author}{Sackett, C.~A.} \emph{et~al.}
\newblock \bibinfo{title}{Experimental entanglement of four particles}.
\newblock \emph{\bibinfo{journal}{Nature}} \textbf{\bibinfo{volume}{404}},
  \bibinfo{pages}{256--259} (\bibinfo{year}{2000}).

\bibitem{Schmidt-Kaler2003}
\bibinfo{author}{Schmidt-Kaler, F.} \emph{et~al.}
\newblock \bibinfo{title}{Realization of the {C}irac-{Z}oller controlled-{NOT}
  quantum gate.}
\newblock \emph{\bibinfo{journal}{Nature}} \textbf{\bibinfo{volume}{422}},
  \bibinfo{pages}{408--411} (\bibinfo{year}{2003}).

\bibitem{Leibfried2003a}
\bibinfo{author}{Leibfried, D.} \emph{et~al.}
\newblock \bibinfo{title}{Experimental demonstration of a robust, high-fidelity
  geometric two ion-qubit phase gate.}
\newblock \emph{\bibinfo{journal}{Nature}} \textbf{\bibinfo{volume}{422}},
  \bibinfo{pages}{412--415} (\bibinfo{year}{2003}).

\bibitem{Haljan2005b}
\bibinfo{author}{Haljan, P.~C.} \emph{et~al.}
\newblock \bibinfo{title}{Entanglement of trapped-ion clock states}.
\newblock \emph{\bibinfo{journal}{Phys. Rev. A}} \textbf{\bibinfo{volume}{72}},
  \bibinfo{pages}{062316} (\bibinfo{year}{2005}).

\bibitem{Home2006}
\bibinfo{author}{Home, J.~P.} \emph{et~al.}
\newblock \bibinfo{title}{Deterministic entanglement and tomography of ion spin
  qubits}.
\newblock \emph{\bibinfo{journal}{New J. Phys.}} \textbf{\bibinfo{volume}{8}},
  \bibinfo{pages}{188} (\bibinfo{year}{2006}).

\bibitem{Riebe:2006a}
\bibinfo{author}{Riebe, M.} \emph{et~al.}
\newblock \bibinfo{title}{Process tomography of ion trap quantum gates}.
\newblock \emph{\bibinfo{journal}{Phys. Rev. Lett.}}
  \textbf{\bibinfo{volume}{97}}, \bibinfo{pages}{220407}
  (\bibinfo{year}{2006}).

\bibitem{Milburn2000}
\bibinfo{author}{Milburn, G.~J.}, \bibinfo{author}{Schneider, S.} \&
  \bibinfo{author}{James, D. F.~V.}
\newblock \bibinfo{title}{Ion trap quantum computing with warm ions}.
\newblock \emph{\bibinfo{journal}{Fortschr. Phys.}}
  \textbf{\bibinfo{volume}{48}}, \bibinfo{pages}{801--810}
  (\bibinfo{year}{2000}).

\bibitem{Solano1999}
\bibinfo{author}{Solano, E.}, \bibinfo{author}{de~Matos~Filho, R.~L.} \&
  \bibinfo{author}{Zagury, N.}
\newblock \bibinfo{title}{Deterministic {B}ell states and measurement of the
  motional state of two trapped ions}.
\newblock \emph{\bibinfo{journal}{Phys. Rev. A}} \textbf{\bibinfo{volume}{59}},
  \bibinfo{pages}{R2539--R2543} (\bibinfo{year}{1999}).

\bibitem{Roos2007}
\bibinfo{author}{Roos, C.~F.}
\newblock \bibinfo{title}{Ion trap quantum gates with amplitude-modulated laser
  beams}.
\newblock \emph{\bibinfo{journal}{New Journal of Physics}}
  \textbf{\bibinfo{volume}{10}}, \bibinfo{pages}{013002}
  (\bibinfo{year}{2008}).

\bibitem{Benhelm2007}
\bibinfo{author}{Benhelm, J.} \emph{et~al.}
\newblock \bibinfo{title}{Measurement of the hyperfine structure of the
  {S}$_{1/2}$-{D}$_{5/2}$ transition in $^{43}${C}a$^+$}.
\newblock \emph{\bibinfo{journal}{Phys. Rev. A}} \textbf{\bibinfo{volume}{75}},
  \bibinfo{pages}{032506} (\bibinfo{year}{2007}).

\bibitem{Roos2006}
\bibinfo{author}{Roos, C.~F.}, \bibinfo{author}{Chwalla, M.},
  \bibinfo{author}{Kim, K.}, \bibinfo{author}{Riebe, M.} \&
  \bibinfo{author}{Blatt, R.}
\newblock \bibinfo{title}{'{D}esigner atoms' for quantum metrology.}
\newblock \emph{\bibinfo{journal}{Nature}} \textbf{\bibinfo{volume}{443}},
  \bibinfo{pages}{316--319} (\bibinfo{year}{2006}).

\bibitem{Moelmer1999}
\bibinfo{author}{M{\o}lmer, K.} \& \bibinfo{author}{S{\o}rensen, A.}
\newblock \bibinfo{title}{Multiparticle entanglement of hot trapped ions.}
\newblock \emph{\bibinfo{journal}{Phys. Rev. Lett.}}
  \textbf{\bibinfo{volume}{82}}, \bibinfo{pages}{1835--1838}
  (\bibinfo{year}{1999}).

\bibitem{Leibfried2005}
\bibinfo{author}{Leibfried, D.} \emph{et~al.}
\newblock \bibinfo{title}{Creation of a six--atom '{S}chr\"odinger cat' state.}
\newblock \emph{\bibinfo{journal}{Nature}} \textbf{\bibinfo{volume}{438}},
  \bibinfo{pages}{639--642} (\bibinfo{year}{2005}).

\bibitem{Chiaverini2004}
\bibinfo{author}{Chiaverini, J.} \emph{et~al.}
\newblock \bibinfo{title}{Realization of quantum error correction.}
\newblock \emph{\bibinfo{journal}{Nature}} \textbf{\bibinfo{volume}{432}},
  \bibinfo{pages}{602--605} (\bibinfo{year}{2004}).

\bibitem{Reichle2006a}
\bibinfo{author}{Reichle, R.} \emph{et~al.}
\newblock \bibinfo{title}{{E}xperimental purification of two-atom
  entanglement.}
\newblock \emph{\bibinfo{journal}{Nature}} \textbf{\bibinfo{volume}{443}},
  \bibinfo{pages}{838--841} (\bibinfo{year}{2006}).

\bibitem{Vandersypen2004}
\bibinfo{author}{Vandersypen, L. M.~K.} \& \bibinfo{author}{Chuang, I.~L.}
\newblock \bibinfo{title}{{NMR} techniques for quantum control and
  computation}.
\newblock \emph{\bibinfo{journal}{Rev. Mod. Phys.}}
  \textbf{\bibinfo{volume}{76}}, \bibinfo{pages}{1037--1069}
  (\bibinfo{year}{2004}).

\bibitem{Haeffner2003}
\bibinfo{author}{H\"affner, H.} \emph{et~al.}
\newblock \bibinfo{title}{Precision measurement and compensation of optical
  stark shifts for an ion-trap quantum processor.}
\newblock \emph{\bibinfo{journal}{Phys. Rev. Lett.}}
  \textbf{\bibinfo{volume}{90}}, \bibinfo{pages}{143602}
  (\bibinfo{year}{2003}).

\bibitem{Leibfried2007}
\bibinfo{author}{Leibfried, D.}, \bibinfo{author}{Knill, E.},
  \bibinfo{author}{Ospelkaus, C.} \& \bibinfo{author}{Wineland, D.~J.}
\newblock \bibinfo{title}{Transport quantum logic gates for trapped ions}.
\newblock \emph{\bibinfo{journal}{Phys. Rev. A}} \textbf{\bibinfo{volume}{76}},
  \bibinfo{pages}{032324} (\bibinfo{year}{2007}).

\end{thebibliography}


\newpage
Addendum
\begin{addendum}
\item[Acknowledgements] We gratefully acknowledge the support of
the European network SCALA and the Disruptive Technology Office
and the Institut f\"ur Quanteninformation GmbH. We thank
R.~Gerritsma and F.~Z{\"a}hringer for help with the experiments.

\item[Competing Interests] The authors declare that they have no
competing financial interests.

\item[Correspondence] Correspondence and requests for materials
should be addressed to C.F.R. \newline
(email:\mbox{Christian.Roos@uibk.ac.at}).

\end{addendum}


\newpage

\newpage

\newpage

\newpage

\end{document}